\documentclass[aps,prx,twocolumn,groupedaddress,nofootinbib]{revtex4-1}

\usepackage{hyperref}
\usepackage{CJK}
\usepackage{amsmath}
\usepackage{graphicx}
\usepackage{dcolumn}
\usepackage{enumerate,amsthm,amssymb,color}
\usepackage{bbm}
\usepackage{bm}
\usepackage[normalem]{ulem}
\usepackage[french, english]{babel}
\usepackage{synttree}
\usepackage{caption}
\usepackage{subfig}
\usepackage{multirow}
\usepackage{xfrac}
\usepackage{fixltx2e}
\usepackage{comment}

\newcommand{\beq}{\begin{equation}}
\newcommand{\eeq}{\end{equation}}
\newcommand{\bea}{\begin{align}}
\newcommand{\eea}{\end{align}}
\newcommand{\bq}{\begin{quote}}
\newcommand{\eq}{\end{quote}}

\newcommand{\blk}{\color{black}}

\begin{document}
\title{Reply to ``Comment on `Why interference phenomena do not capture the essence of quantum theory'\! ''}
\author{Lorenzo Catani}\email{lorenzo.catani@tu-berlin.de}
\affiliation{Electrical Engineering and Computer Science Department, Technische Universit\"{a}t Berlin, 10587 Berlin, Germany}

\author{Matthew Leifer}\email{leifer@chapman.edu}\affiliation{Institute for Quantum Studies and Schmid College of Science and Technology, Chapman University, One University Drive, Orange, CA, 92866, USA}

\author{David Schmid}\email{david.schmid@ug.edu.pl}\affiliation{International Centre for Theory of Quantum Technologies, University of Gdansk, 80-308 Gdansk, Poland}

\author{Robert W. Spekkens}\email{rspekkens@perimeterinstitute.ca}
\affiliation{Perimeter Institute for Theoretical Physics, 31 Caroline Street North, Waterloo, Ontario Canada N2L 2Y5}


\begin{abstract}
Our article~\cite{catani2021interference} argues  that  the phenomenology of interference that is traditionally regarded as problematic does {\em not}, in fact, capture the essence of quantum theory---contrary to the claims of Feynman and many others.  It does so by demonstrating the existence of  a physical theory, which we term the ``toy field theory'', that reproduces this phenomenology but which does not sacrifice the classical worldview.  In their Comment~\cite{HanceHossenfelderComment}, Hance and Hossenfelder dispute our claim.
Correcting mistaken claims found therein and responding to their criticisms provides us with an opportunity to further clarify some of the ideas in our article. 
\end{abstract}

\maketitle




\section{On epistemic states and epistemic restrictions}

Hance and Hossenfelder 
 state 
\begin{quote}
In Spekkens' original toy model paper [6], an ``ontic state'' is defined as
``a state of reality'' whereas an ``epistemic state'' is a ``state of knowledge''. Both of these
definitions are useless, the first because one does not know what ``reality'' means, the
second because one does not know what ``knowledge'' means [...] 
\end{quote}

At some level, no one can reasonably claim that they fail to understand the distinction between reality and our knowledge thereof. We all understand the difference between the proposition ``the back door is locked'' and the proposition ``I know that the back door is locked''.  Hance and Hossenfelder are presumably not suggesting that they fail to comprehend the distinction in this common-sense form.  Rather, they are presumably suggesting that the distinction introduced in Ref.~\cite{Spekkens2007} is deficient by virtue of not having been adequately formalized. 

Such a criticism might have been apt if all Ref.~\cite{Spekkens2007} had to offer in the way of trying to clarify the distinction between ontic states and epistemic states was an explanation of why it saw fit to use this terminology, namely, that the former term derives from the Greek {\em ontos}, meaning ``to be'', and the latter from the Greek {\em episteme}, meaning ``knowledge''. As it turns out, however, the discussion 
of this distinction in Ref.~\cite{Spekkens2007} did not, in fact, end with an explanation of the etymology.  Indeed, immediately after introducing this terminology, it is stated that ``To understand the content of the distinction, it is useful to study how it arises in the uncontroversial context of classical physics,'' followed by this elucidation of the concept:
\begin{quote}
The first notion of state that students typically encounter in their study of classical physics is the one associated with a point in phase space. This state provides a complete specification of all the properties of the system---in particle mechanics, such a state is sometimes called a ``Newtonian state''. It is an ontic state. On the other hand, when a student learns classical statistical mechanics, a new kind of state is introduced, corresponding to a probability distribution over the phase space---sometimes called a ``Liouville state''. This is an epistemic state. The critical difference between a point in phase space and a probability distribution over phase space is not that the latter is a function. An electromagnetic field configuration is a function over three-dimensional space, but is nonetheless an ontic state. What is critical about a probability distribution is that the relative height of the function at two different points is not a property of the system---unlike the relative height of an electromagnetic field at two points in space. Rather, this relative height represents the relative likelihood that some agent assigns to the two ontic states associated with those points of the phase space. The distribution describes only what this agent knows about the system.
\end{quote}
In other words, the distinction between ontic states and epistemic states is  not novel to Ref.~\cite{Spekkens2007}.  It is already present in physics whenever a system being investigated is such that the investigator may have incomplete knowledge of its physical state.  Classical statistical mechanics is the field of physics wherein it first became critical to develop a mathematical formalism for describing such incomplete knowledge.  
Thus, for instance, a microcanonical ensemble for a gas represents a state of incomplete knowledge, appropriate for any agent who knows only certain macroscopic properties of the gas.

Moreover, Ref.~\cite{Spekkens2007} goes on to formalize the distinction in terms of a discrete ontic state space and the space of probability distributions thereon.  The same was done for continuous ontic state spaces in Ref.~\cite{Bartlett2012}, where epistemic states are now probability densities over the ontic state space. 

More generally, the ontic states of a system are the elements of the set that defines the kinematics for the system, and where functions from this set to itself define the dynamical laws.  Epistemic states, on the other hand, belong to the normative theory for reasoning in the face of uncertainty, such as Bayesian probability theory.  They describe the elements of the set of possible ways of knowing about the ontic state of the system according to the theory.
A synthetic approach to the distinction, aiming to go beyond the classical case, has recently been presented in Ref.~\cite{Schmid2021unscrambling}.

As Hance and Hossenfelder do not comment on any of the formal accounts of the distinction between ontic and epistemic states that we have just outlined, their criticism of the distinction  seems to us to be devoid of any substance. 


Hance and Hossenfelder ask two further questions about the status of epistemic states: ``{\em whose} knowledge?''
 and ``why would we care  about it?''

Take the second question first.  
The importance of mathematically formalizing our uncertainty is  apparent in all branches of the sciences, including physics.
It arises in every situation where there is incomplete knowledge due to practical considerations, such as technological limitations.  
If the incomplete knowledge is due to a fundamental feature of the physical theory being considered rather than a technological limitation---as with the epistemic restriction in Ref.~\cite{Spekkens2007} and our toy field theory---this does not alleviate the need to formally quantify uncertainty.  Rather, it makes it more acute.



 

What about the question ``whose knowledge?''  Note, first of all,  that one could ask Hance and Hossenfelder's question about states of incomplete knowledge in classical statistical mechanics.
Ought we to take the microcanonical ensemble to be an unprofessionally vague concept because it has not been made explicit in the textbooks {\em who it is} that knows the values of certain macrovariables while remaining ignorant of the values of the microvariables?  No, of course not.  The microcanonical ensemble describes the knowledge of {\em any agent} that knows only the specified macro-variables.   The statistical mechanics textbooks are right not to waste space answering the question ``whose knowledge?'' when they ask us to imagine that only certain macroscopic variables are known.

More generally, there is, in fact, a long tradition of expressing physical laws in a pragmatic way, that is, in terms of in-principle restrictions on what any agent living in a universe following those laws 
might be able to do or to know. (See, for instance, the introduction of Ref.~\cite{chiribella2016quantum}.)  
Consider Kelvin's formulation of the second law of thermodynamics.  ``It is impossible to devise a cyclically operating device, the sole effect of which is to absorb energy in the form of heat from a single thermal reservoir and to deliver an equivalent amount of work.'' \cite{rao1997chemical} We might call this principle a ``pragmatic restriction'', to parallel the notion of an epistemic restriction.  Suppose someone asks: 
``Who does this pragmatic restriction apply to?'' or simply ``Who is the `deviser'?''  Is this a question that needs to be answered to understand Kelvin's formulation?  
No.  It is understood that in Kelvin's formulation, the answer to the ``who'' question is {\em any agent at all, using any physically realizable technology whatsoever.}  It is not a parochial kind of restriction, specific to some moment in technological history or some particular engineer.  It is an in-principle kind of restriction.  It is the same with the epistemic restriction in physical theories that posit one. 

Hance and Hossenfelder claim that the epistemic restriction used in Ref.~\cite{catani2021interference} is different from the one used in Ref.~\cite{Spekkens2007}, on the grounds that the one stated in Ref.~\cite{Spekkens2007} is such that the update rule that it implies does not act locally.  This is also mistaken.  There is no contrast between the nature of the epistemic restriction (and consequently the update rules for epistemic states) in Ref.~\cite{catani2021interference} and Ref.~\cite{Spekkens2007}. In particular, both are explicitly local.  The fact that the latter satisfies locality is emphasized throughout Ref.~\cite{Spekkens2007}, in particular, as the reason we know that the toy theory cannot violate Bell inequalities.  It is unclear, therefore, how Hance and Hossenfelder came to this mistaken impression.

We turn to considering the following two claims  of Hance and Hossenfelder
\begin{quote}
After all, we use quantum mechanics to predict frequencies of occurrence and not Peter Pan's knowledge about these frequencies. 
\end{quote}
and
\begin{quote}
Indeed, one may wonder, why talk about knowledge at all? What we need to predict measurement outcomes is a prescription for the distribution of an ensemble of ontic states [5]. The claim of Catani et al that they can correctly reproduce observations only make sense if the ``epistemic restriction'' is a change
to the underlying distribution of ontic states.
\end{quote}



In these quotations,  Hance and Hossenfelder are attacking the idea that probabilities ought to be defined as credences (i.e., an agent's degrees of belief) and seem to instead endorse the notion that they ought to be defined as relative frequencies.  The literature on the philosophy of probability provides many arguments against this type of frequentist interpretation of probability.   
Indeed, it is a rare instance of something about which there seems to be {\em agreement} among those writing on the philosophy of probability. Myrvold, in his recent book on the philosophy of probability~\cite{myrvold2021beyond}, goes so far as to call this view the ``dead horse'' of the philosophy of probability.~\footnote{Note, however, that there are versions of frequentism that do not seek to {\em define} probabilities as relative frequencies~\cite{von1981probability,Gillies2000-GILPTO-2}.}

In our view, the key argument against interpreting probabilities as relative frequencies is the following one.  Relative frequencies connect with the probability distribution assigned to a single run through the law of large numbers.  But what this law states is that, in the limit of infinitely large ensembles, these relative frequencies are {\em likely} to converge to the probabilities in the probability distributions, in the sense that this will occur in a set of measure one of possible sequences. 
For a Bayesian, the notion of ``likelihood'' in the law of large numbers (more formally, the notion of a measure) is an appeal to probability that, like all appeals to probability, ought to be interpreted as a credence. But let us consider the frequentist alternative, that this probability also is to be interpreted as a relative frequency.  This means that one must interpret the law of large numbers as stating that if one forms an infinite ensemble of copies of the original infinite ensemble, the relative frequency with which the convergence occurs in this new ensemble goes to 1.  But it is not the case that the convergence must occur in {\em every} element of the new ensemble.  So, strictly speaking, all one can claim is that in any given element of this new ensemble, it is ``likely'' that the convergence occurs.  But now one is faced with the problem of how to interpret {\em this} notion of likelihood. One can define a third type of ensemble of copies of the second type of ensemble, but then another notion of likelihood appears at that level which needs to be defined.  No matter how far one goes in this sequence, there always remains a concept of probability that remains undefined.  In short, attempts to define probabilities as relative frequencies lead to an infinite regress. \blk

In an appendix that we have added to our article (Appendix C.1),
we discuss at length the question of what the toy field theory has to say about relative frequencies.
 We point out that the connection to relative frequencies comes when considering repetitions of an experiment with an i.i.d.~source.  If ${\bf p}$ is the probability distribution assigned to a system by an agent (i.e., representing the agent's credences about the system), then ${\bf p}^{\otimes n}$ is the probability distribution assigned to the $n$ copies of the system in the $n$-fold repetition of the i.i.d.~experiment (i.e., representing the agent's credences about the $n$ system of the i.i.d. source). 
Then, the law of large numbers stipulates that in the limit of arbitrarily many repetitions of the experiment, the relative frequencies judged to be most likely are those that 
 converge to the distribution ${\bf p}$. 



 
That said, one {\em can} provide an account of our toy field theory in a language that is more congenial to those who are inclined to a frequentist interpretation of probability.   
It suffices to express facts about an infinite ensemble of repetitions of an experiment in terms of the relative frequencies that are likely to occur in this ensemble (without {\em defining} the probabilities in terms of these relative frequencies).  This provides an alternative to the description we provided in the main text of our article (which was in terms of an agent's {\em state of knowledge about a particular element} of this ensemble).   We make this translation in another appendix that we have added to our article (Appendix C.2). 

In this new description, the transformation to the physical state induced by a beamsplitter or phase shifter describes a deterministic change in the make-up of the ensemble of physical states at a particular point in the interferometer (rather than a deterministic change in an agent's state of knowledge about a particular element of the ensemble).
Similarly, in this new description, conditioning on the outcome of a measurement is no longer modelled as Bayesian updating, but as updating the ensemble that is relevant for making predictions.  We describe it as follows in our Appendix C.2:
 \begin{quote}
 This updating can be understood as consisting of two steps. First, one selects from the pre-existing ensemble the subensemble that is consistent with the outcome that was observed. Second, the fact that a measurement leads to a random disturbance---that is, one of several different transformations to the physical state---implies that elements of the ensemble selected in the first step get split into distinct elements (bifurcated in the case of interest here), thereby leading to an increase in the number of distinct subensembles.\end{quote}
 For the case of a measurement on one mode, the ensemble of possibilities for the physical state of the other mode may be updated because of a pre-existing correlation between the physical states of the two modes.  Consequently, such updating does not involve any nonlocal influence.  
 
\color{black} 
 
Hance and Hossenfelder also state:
\begin{quote}
To further muddy the waters, by virtue of the ``epistemic restriction'' of Spekkens' original toy model, no observer can ``know'' what the ``ontic'' state is, which makes it rather unclear what it might mean for it to be ``real'' in the first place.
\end{quote}

If one grants that there is a meaningful distinction between ontic states and epistemic states, 
then it is perfectly straightforward to imagine a scenario wherein agents can have knowledge of any of a number of aspects of the full reality, while not being able to know {\em all} of these aspects at once.  

Plato's allegory of the cave is a useful metaphor for this aspect of epistemically restricted theories~\cite{spekkens2015paradigm}.  One can imagine that the objects that are casting the shadows have a three-dimensional shape and the prisoners in the cave, by virtue of only seeing the shadows, can only come to know various two-dimensional projections of these three-dimensional shapes.  Indeed, we might imagine that a given shape can be oriented in an arbitrary way relative to the light source at the mouth of the cave, such that the prisoners can come to learn {\em any} two-dimensional projection, while still never being able to access more than a single two-dimensional projection at a time.  This limitation of the prisoners, to only ever acquiring partial information about the shape at a given time, does not in any way undermine the notion that there is, in fact, a three-dimensional shape of which the shadow they see is a two-dimensional projection. 


The only way we see for someone to think that it {\em does} undermine this notion is if they subscribe to
the {\em verificationist principle} of the logical positivists.\footnote{Logical positivism is a type of empiricism (or instrumentalism) that was popular in the early twentieth century.}
  This is the idea that a proposition is only meaningful if it is possible to conduct an experiment that verifies it.  
For the prisoners in Plato's cave, there is no experimental procedure by which they can come to learn all of the two-dimensional projections of an object at once, that is, its three-dimensional shape.  It follows that endorsement of the verificationist principle stipulates  that it is not meaningful to talk about its three-dimensional shape.   Similarly, given that in an epistemically restricted theory there is no way to simultaneously measure all the ontic variables (i.e., every variable in a set that is sufficient to determine the ontic state), endorsement of the verification principle implies endorsement of the notion that propositions about the values of such a set of ontic variables are not jointly meaningful.


This sort of criticism of epistemically restricted theories is reminiscent of Bohr's position in the Bohr-Einstein debate.  In Ref.~\cite{Bartlett2012}, it was argued that Bohr's account of experiments measuring alternatively position or momentum of a particle, and in particular his account of the Einstein-Podolsky-Rosen (EPR) experiment, harmonize quite well with the account given by the theory described therein---termed epistemically restricted Liouville  mechanics (abbreviated as ERL mechanics)---where a particle has both a position and a momentum, but these can never be known simultaneously.  It was furthermore argued in Ref.~\cite{Bartlett2012} that the reason Bohr ultimately rejects this account is that he endorsed a version of the verificationist principle.  The argument was as follows:
\begin{quote}
[...] ERL mechanics can reproduce the correlations in the original EPR thought experiment and indeed delivers the sort of interpretation of the correlations that EPR favoured, namely, one wherein position and momentum are jointly well-defined but not jointly known. Even though Bohr sought to dispute this sort of interpretation in his reply, his description of the thought experiment makes explicit reference to the positions and momenta of the systems: ``In fact, even if we knew the position of the diaphragm relative to the space frame before the first measurement of its momentum, and even though its position after the last measurement can be accurately fixed, we lose, on account of the uncontrollable displacement of the diaphragm during each collision process with the test bodies, the knowledge of its position when the particle passed through the slit.'' Indeed, his argument for the consistency of the uncertainty principle makes no reference to the quantum formalism at all. It reads better as an argument for the consistency of the uncertainty principle within ERL mechanics. Nonetheless, Bohr denies the interpretation suggested by ERL mechanics: ``we have in each experimental arrangement suited for the study of proper quantum phenomena not merely to do with an ignorance of the value of certain physical quantities, but with the impossibility of defining these quantities in an unambiguous way.'' The only way we see to reconcile this tension in Bohr's reply is that Bohr believed that two quantities can be jointly well-defined {\em only} if they can be jointly measured. In essence, Bohr was a radical positivist. Otherwise, why from the impossibility of two quantities being jointly measured would he infer the impossibility of their being jointly {\em well-defined}, as opposed to merely inferring the impossibility of their being jointly {\em known}?
\end{quote}

More generally, there have been many persuasive arguments put forward {\em against} the verificationist principle  and the positivist movement in the philosophy of science more generally.
For those not familiar with these arguments, we recommend Quine's classic article ``Two dogmas of empiricism''~\cite{quine1976two}.
 
\section{Classicality}

In Sec.~V.A.3 of our article, when considering what might be genuinely nonclassical about interference phenomenology, we describe our own preferred notion of classicality.  Hance and Hossenfelder's comment includes some criticisms of this notion.
It should be noted, however, that the thesis of our article was not predicated on our readers espousing the notion of classicality that we favour.  As such, these criticisms are not relevant to the question of primary interest in our article.   Nonetheless, we take this opportunity to respond to them for the sake of clarifying what our preferred notion of classicality implies.\blk


 
Hance and Hossenfelder correctly summarize our preferred notion of classicality, which includes a notion of classicality for the theory of inference, namely, that it is done using
Bayesian probability theory and Boolean propositional logic.  Nonetheless, they suggest that according to our preferred notion, what is usually termed ``classical statistical mechanics'' would come out as nonclassical.  This is incorrect.  What they seem to have missed is that the contrast class we had in mind in our notion, the thing that we {\em would} call ``nonclassical'', is a theory wherein the way inferences are done is {\em at odds with} Bayesian probability theory and Boolean propositional logic.    

The use of probability in statistical mechanics is not at odds with either Bayesian probability theory or Boolean propositional logic.  Consequently, there is no sense in which statistical mechanics would come out as nonclassical according to our preferred notion of classicality.  
Indeed, Hance and Hossenfelder do not suggest that classical statistical mechanics uses some exotic {\em alternative} to Bayesian probability theory, but only that ``Bayesian probability theory isn't used much'' in classical statistical mechanics.  Although we could dispute this assessment of the prevalence of Bayesian inference in classical statistical mechanics (see, e.g., Ref.~\cite{jaynes1957information}), it is beside the point. 
 {\em Even if} for some physical theory, Bayesian probability theory {\em was not used at all}, this would not imply that that physical theory would be judged to be nonclassical relative to our notion.  A physical theory needs to commit itself to some {\em concrete alternative} to Bayesian probability theory or Boolean propositional logic 
  in order for it to be judged as having a nonclassical theory of inference by the lights of our preferred notion.  Lack of use is not the same as use of an alternative.
  
It is worth adding that we do not consider modifications to the {\em interpretation} of probabilities (without any difference to the formal apparatus for making predictions) to be examples of a concrete alternative to the classical theory of inference.  Such an alternative must deviate from Bayesian probability theory and Boolean propositional logic in more than a cosmetic manner.    As such, a mere preference for understanding the predictions of classical statistical mechanics in terms of a frequentist interpretation of probabilities rather than a Bayesian one is not sufficient for claiming that the theory of inference used therein is nonclassical.  Similarly, a modification to the {\em scope} of some theory of inference, such as moving from Boolean propositional logic to standard predicate logic also does not constitute an example of a concrete alternative to Boolean propositional logic since the propositional segment of predicate logic is still Boolean.
Examples of the sorts of modifications of logic that we {\em would} consider to be at odds with Boolean propositional logic are {\em quantum logics}~\cite{hooker1978logicoalgebraic}.  Similarly, an example of a modification of probability theory that we would consider to be at odds with Bayesian probability theory is the sort of theory of inference defined in Ref.~\cite{LeiferSpekkens,coecke2012picturing,horsman2017can} using conditional density operators.\blk 

\blk


Hance and Hossenfelder also seek to criticize the Leibnizian methodological principle that is part of our preferred notion of classicality.  They state:
\begin{quote}
What is empirically `indiscernible' depends on what measurements one has made or can make. Distances below, say, a thousandth of a femtometer aren't currently `empirically discernible'. We treat them as ontologically different in General Relativity, hence, it seems that according to the authors' position, General Relativity is not a classical theory.
\end{quote}
Here, Hance and Hossenfelder are simply mistaken about the content of the Leibnizian methodological principle.  The definition from Ref.~\cite{spekkens2019ontological}, which was repeated in Ref.~\cite{catani2021interference}, states that the notion of empirical discernibility at issue is indistinguishability {\em in principle} rather than in practice, where what is possible in principle is determined by the physical theory that one is assessing.

The point is emphasized in Ref.~\cite{spekkens2019ontological}:
\begin{quote}
[...] the Leibnizian methodological principle does not appeal to a parochial kind of empirical indiscernibility, judged relative to the particular in-born capabilities of humans or their particular technological capabilities at a given historical moment, but rather to the in-principle variety of empirical indiscernibility. This variety of indiscernibility must be understood as indiscernibility for any system that might be considered an agent within the universe. This is because, as Deutsch has argued persuasively, the only in-principle limits to human capabilities are the limits imposed by physics [*], and therefore the only limits on our capabilities are the limits on the capabilities of any system embedded in the universe and subject to its physical laws.  [*] His argument proceeds by noting that an ``in-principle human capability'' includes what could be achieved in a distant future with the aid of arbitrarily sophisticated technology.
\end{quote}

Thus, the fact that certain distances that are not distinguishable by today's technology are nonetheless treated as ontologically distinct in General Relativity is not a failure of the Leibnizian methodological principle.  Only if General Relativity stipulated that such distances were {\em in principle} empirically indistinguishable, would one conclude that General Relativity contradicted the principle.   In fact, the Leibnizian methodological principle is built into General Relativity at a deep level, since this is one of main principles that guided Einstein in his development of the theory, 
 as is argued in Ref.~\cite{spekkens2019ontological}.

Although our article did not seek to persuade readers to endorse our preferred notion of classicality, it {\em did} seek to insist 
on a methodological point, namely, that if someone wants to claim that some particular operational phenomenology of interference {\em does} capture the essence of quantum theory, then  they ought to back up their view with a no-go result.  That is, they should articulate a formal notion of classicality within some framework for physical theories and then prove a theorem demonstrating that their notion is inconsistent with the phenomenology in question.  
It is important to stipulate a formal notion of classicality in such a no-go result because we do not, in fact, 
 all agree about what the correct notion of classicality is.  
 Indeed, there are almost as many ideas about this as there are researchers who work in the foundations of quantum theory. By abiding by the proposed methodology, one can focus the discussion on where the true disagreements lie.  This is discussed further in the next section.


\section{On shifting the goal posts}

Hance and Hossenfelder attempt to summarize part of how our theory works as follows:
\begin{quote}
The phase of the state changes when a measurement doesn't happen, [...] 
\end{quote}
and then seek to critique it based on this characterization:
\begin{quote}
[...] it is unclear how the absence of a measurement can locally change a state.  
\end{quote}
The summary is incorrect, however. 
 In our theory, any mode for which no measurement is performed has its phase left invariant.  
 Only if a measurement is actually performed on a mode can the latter's phase be randomized.    
 

Hance and Hossenfelder also state:
\begin{quote}
The authors seem to assume that a measurement in which no interaction happens still somehow results in an interaction (and that this interaction is still local).
\end{quote}
Here, they at least seem to acknowledge that the measurement update rule we describe in our article {\em applies to the case where a measurement of occupation number is happening}, as opposed to no measurement happening.  However, their  claim that no interaction happens as a result of this measurement is mistaken.  
 In our theory, every mode has a phase degree of freedom in addition to its occupation number degree of freedom, and the phase of a mode is randomized in a measurement of occupation number of that mode {\em regardless of whether the occupation number is found to be 0 or 1},
 i.e., regardless whether or not the excitation happens to be found in that mode.  
It is likely that Hance and Hossenfelder's confusion results from thinking of our theory as one wherein the systems are particles, when in fact the systems in our theory are modes.  To head off such confusions, we have clarified this distinction in two appendices we have added to our article (Appendix C.5 and C.6).

Hance and Hossenfelder correctly summarize an aspect of our theory  when they note that it posits that information is sent over a path with occupation number zero.  They are again mistaken, however, when they state that 
\begin{quote}
This only works so long as one is forbidden from blocking one of the paths or pulling out one of the mirrors. On doing either of these things, the model either falls apart or requires nonlocal update.
\end{quote}
 In two more appendices we have added to our article (Appendices C.3 and C.4), we have provided an explicit treatment of each of these cases 
and demonstrated that the theory has no problem treating either.  It reproduces what one would expect for the analogous quantum experiment and continues to only require local causal influences to do so.  

As we note in the new Appendix C.3, blocking a path amounts to implementing a destructive measurement of occupation number, one that absorbs the excitation if it is present.  Consequently, one way to summarize Hance and Hossenfelder's first claim is that the toy field theory cannot account for the Mach-Zehnder interferometer experiment in the case where the measurement of the occupation number is made to be destructive rather than nondestructive.



They correctly anticipate that our response to this suggestion is that it is an instance of what we called ``shifting the goal posts'' in our article.\footnote{They suggest further that we ourselves are guilty of some kind of shifting of the goal posts in our original paper by treating the Mach Zehnder experiment rather than the double-slit experiment.  Our move is not an instance of the specific argumentative strategy we called ``shifting the goal posts'' in our article (which we describe below), so it is not really a tit-for-tat situation, as Hance and Hossenfelder suggest.  Nonetheless, we here respond to the charge that we have done less than we needed to do to justify the thesis of our article.  In the introduction of our article, we provided a summary of the arguments put forward by Feynman and others in favour of the three interpretational claims and the impossibility of a classical explanation {\em within the context of the double-slit experiment}.  Then, in Sec. II.A, we show that the argument, {\em when adapted to 
 the context of the Mach-Zehnder experiment}, is of precisely the same form.  In other words, the specific aspects of the operational phenomenology that are  cited in the argument
   are {\em common} to the double-slit and Mach-Zehnder scenarios, as is the logical form of the argument. 
As such, showing that the claimed implication is invalid in any particular scenario shows that the argument is {\em not valid} (in the logician's sense of the conclusion failing to follow from the premisses), and so is not to be trusted in any context in which it arises. It is for this reason that
undermining the argument in the context of the Mach Zehnder experiment is sufficient to undermine it also in the context of the double-slit experiment.  }

We begin by recalling how this notion 
was articulated in Ref.~\cite{catani2021interference}:
\begin{quote}
No doubt those researchers who are sympathetic to the view that interference captures the essence of quantum theory will be tempted to respond to the arguments of this article as follows: ``Sure, you have reproduced {\em some} of the phenomenology of quantum interference, but you haven't reproduced {\em all} of it. What about all of the experiments involving beamsplitters that are not 50-50, or involving phase shifts other than $\Phi =0$ and $\Phi = \pi$? You can't make sense of {\em those} in the toy field theory.
\end{quote}
Does Hance and Hossenfelder's response fit the pattern described here?  Yes, it does. Indeed, their response is essentially this: ``Sure, you might have reproduced the phenomenology of quantum interference in the case where the measurement of occupation number on one of the modes is {\em nondestructive}, but what about the case where the measurement is {\em destructive}?''

As we note in our new appendix, the thesis of our article does not rely on addressing this case.
\begin{quote}
Explicit in Feynman's account (and to an even greater degree in Elitzur and Vaidman's account) is that the destruction of interference occurs even in the case where the measurement {\em does not detect the excitation on its arm}.  But in this case, {\em there is no difference} in the quantum state update rule between destructive and nondestructive measurements, as the output of the measurement device is left in the quantum vacuum state in both cases.  For this reason, the distinction between destructive and nondestructive measurements 
is not significant for discussions of the TRAP phenomenology.
 It follows that if one can reproduce this phenomenology for {\em either} type of measurement in a classical local model, one has undermined the claim that the phenomenology necessitates a departure from the classical worldview. Hence, an explicit consideration of destructive measurements is not required to establish our thesis.
\end{quote}
(Here, ``TRAP'' is an abbreviation of `traditionally regarded as problematic'.)
 The same point can be made regarding any 
 modification of the experimental scenario wherein a mirror is removed, since this case appears nowhere in discussions of what is typically regarded as problematic about interference in quantum theory. 


It is odd that while Hance and Hossenfelder anticipated that we would identify their question as an instance of what we termed ``shifting the goal-post'', they did not bother to consider what we said about how best to approach such questions.
 This is what we wrote:
\begin{quote}
 If someone wishes to claim that aspects of interference beyond the TRAP phenomenology demonstrate the impossibility of maintaining a classical worldview, then not only must they specify precisely which aspects they have in mind and how they propose to formalize the notion of classicality,  they must also {\em back up their claim} with a rigorous no-go theorem, following the methodology we endorsed above. Until they do, the view that the phenomena in question resists explanation in terms of a classical worldview is mere speculation, and might only indicate a ``lack of imagination'', to recall Bell's phrase.
\end{quote}
It seems to us that Hance and Hossenfelder {\em do} wish to claim that aspects of interference beyond the TRAP phenomenology demonstrate the impossibility of maintaining a classical worldview.  
However, they fail to articulate the precise notion of classicality they have in mind, nor the precise set of  operational features of quantum theory that they are appealing to, nor do they prove a no-go theorem to back up their claim.
  As such, their view, namely, that shifting attention to destructive measurements (or cases where a mirror is removed) implies that the interference phenomena resist explanation in terms of a classical worldview, is {\em mere speculation} and as a result it might merely indicate a lack of imagination on their part for what such a classical explanation might be.  

Indeed, this is precisely what we show to be the case in the appendices we have added to our article (Appendices C.3 and C.4).  

In the conclusions of Ref.~\cite{catani2021interference}, we noted that Feynman could have avoided making the mistaken claim that the  phenomenology of interference resisted explanation in terms of a classical worldview if he had 
tried (and necessarily failed) to back up his belief with a no-go theorem. The same can be said of  Hance and Hossenfelder's mistaken claim regarding the phenomenology of interference with destructive measurements rather than nondestructive measurements, or with a mirror removed.  In this sense, their claim provides an illustration of why one ought to follow the methodology we proposed in our article.
We therefore take this opportunity to repeat a maxim expressed in the introduction of our article:
\begin{quote}
One should not be credulous of statements that a given operational phenomenology implies some interpretational claim unless the statement is backed up by a rigorous no-go theorem proving the implication (typically against the backdrop of additional assumptions).
\end{quote}
We believe that broader adherence to this maxim can 
 raise the level of quality of 
 discussions concerning the foundations of quantum theory particularly 
 between researchers who have diverging interpretational persuasions.

\bibliographystyle{apsrev4-1}
\bibliography{ReplytoHH}

\end{document}